\documentclass[doublecol]{epl2}
\usepackage{graphicx}
\usepackage{amsmath}
\usepackage{makeidx}
\usepackage{indentfirst}
\title{ Covariant Renormalizable Anisotropic Theories and Off-Diagonal
Einstein-Yang-Mills-Higgs Solutions}
\shorttitle{Covariant Renormalizable Anisotropic Theories and Off-Diagonal
EYMH Solutions}

\author{Sergiu I. Vacaru\inst{1}}
\shortauthor{Sergiu I. Vacaru}

\institute{
  \inst{1} Science Department, University "Al. I. Cuza" Ia\c si, 54 Lascar Catargi street, 700107, Ia\c si, Romania
}
\pacs{04.60.-m} { First pacs description} \pacs{04.50.Kd} { Second pacs description}
\pacs{04.20.Jb} { Third pacs description}

\abstract{
We use an important decoupling property of gravitational field equations in the general relativity theory and modifications, written with respect to
nonholonomic frames with 2+2 spacetime decomposition. This allows us to integrate the Einstein equations (eqs) in very general forms with generic off--diagonal metrics depending on all spacetime coordinates via generating and integration functions containing (un--)broken symmetry parameters. We associate families of off-diagonal Einstein manifolds to certain classes of covariant gravity theories which have a nice ultraviolet behavior and seem to be (super) renormalizable in a sense of covariant modifications of Ho\v rava-Lifshits gravity. The apparent
breaking of Lorentz invariance is present in some "partner" anisotropically
induced theories due to nonlinear coupling  with effective parametric interactions determined by nonholonomic constraints and generic off-diagonal gravitational and matter fields configurations. Finally, we show how the constructions can be extended to include  exact solutions  for conjectured  covariant reonormalizable models with  Einstein-Yang-Mills-Higgs fields.
}
\begin{document}

\maketitle

\section{Introduction}
 One of the main difficulties in elaborating viable
models of quantum gravity (QG) is that perturbations in the general
relativity theory (GR) from a flat Minkowski spacetime result in
non--renormalizable divergences from the ultraviolet region in momentum
space. This problem is caused by the dimension of gravitational (Newton)
constant and seems impossible to be solved for a unitary theory with four
fundamental interactions which has Lorentz--invariance even some
higher--derivative,  spin--foam, loop type etc models of gravity may be renormalizable, see
\cite{alvarez,buch,oriti}
for reviews and references. A
recent new approach to QG is based on idea \cite{horava} to construct
Lorentz non--invariant theories with scaling properties of space, $x^{i},$
and time, $t,$ coordinates considered in the form $(\mathbf{x},t)\rightarrow
(b\mathbf{x},b^{z}t),$ where $z=2,3,...$ In such cases, the ultraviolet (UV)
behavior of the graviton propagator changes as $1/|\mathbf{k}%
|^{2}\rightarrow 1/|\mathbf{k}|^{2z},$ where $\mathbf{k}$ is the spacial
momenta. It is possible to elaborate models of QG which are UV
renormalizable, for instance, for $z=3$ but with the price of introducing
some terms breaking the Lorentz invariance explicitly. \ Due to the lack of
full diffeomorphysm invariance and, for instance, impossibility to exclude
completely certain unphysical modes, such theories where criticized in a
series of works (see, for instance, \cite{lumei,moffat}).

In \cite{odints1},
a covariant renormalizable gravity model
developing the Ho\v rava--like gravity to full diffeomorphysm invariance was elaborated. In such an approach, the Lorentz--invariance  of the
graviton propagator is broken via a non--standard coupling with an unknown fluid. Some versions of such theories seem to be (super--) renormalizable, when physical transverse modes may appear. Certain applications in modern cosmology, with accelerating solutions and possible power law inflationary stage, seem to be of substantial interest.

The results of research in \cite{ijgmmp2,vexol} conclude that the
gravitational field eqs in GR and various modifications can be
decoupled with respect to certain classes of nonholonomic frames. It is
possible to integrate the system of Einstein--Yang--Mills and Higgs (EYMH) eqs in very general forms. For such solutions, the generic
off--diagonal metrics can not be diagonalized via coordinate transforms and depend on all spacetime coordinates via corresponding
generating and integration functions and possible (broken, or preserving)
symmetry parameters. Choosing necessary types of such functions and
parameters, via corresponding deformations of frame, metric and connection structures,  and imposing non--integrable (nonholonomic) transforms, we can model effective nonlinear interactions, with scaling properties and local anisotropies, which can be  renormalizable.

We start our constructions with a set of actions $\ ^{[i]}S=\frac{1}{\kappa
^{2}}\int d^{4}u\sqrt{|\mathbf{g}|}\ ^{[i]}\mathcal{L}$ on a four
dimensional (4-d) pseudo--Riemannian manifold $V$ for  Lagrangian
\begin{eqnarray}
\ ^{[1]}\mathcal{L}&=&\ ^{s}\widehat{R}-\widehat{\Lambda }=~^{[2]}\mathcal{L}%
=~^{s}\widetilde{R}+\widetilde{L}=\ ^{[3]}\mathcal{L} \notag \\
&=&R+L(T^{\mu \nu
},R_{\mu \nu })=~^{[4]}\mathcal{L}=F\left( \breve{R}\right) .  \label{act1}
\end{eqnarray}%
In our works, we  use left up and low labels to geometric/physical
objects which may have the same coefficients in a system of reference but
subjected to different types of eqs for certain analogous/effective
theories. All terms in such formulas are stated by the same metric structure
$\mathbf{g}=\{g_{\alpha \beta }\}$ for standard and/or modified models of
gravity theory generated by different (effective) Lagrangians when curvature
scalars $(\ ^{s}\widehat{R},\ ^{s}\widetilde{R},R,\breve{R}),$ cosmological
constant $\widehat{\Lambda },$ non--standard coupling of Ricci, $R_{\mu \nu
},$ and energy--momentum, $T^{\mu \nu },$ tensors via $L(...)$ and $F\left(
...\right) $ etc. The above values and other possible constants and terms
are such way postulated that $\ ^{[1]}\mathcal{L}$ is used for
constructing general classes of generic off--diagonal solutions in GR
  \cite{ijgmmp2,vexol}  but $\ ^{[3]}\mathcal{L}$ is of necessary type to get
a covariant renormalization gravity as in
\cite{odints1}.
For corresponding conditions on $F$ the value $\ ^{[4]}\mathcal{L}$ is related to modified theories \cite{kluson}.
To study  QG models, the spacetime metric  for
theories derived from $\ ^{[3]}\mathcal{L}$ and/or $\ ^{[4]}\mathcal{L}$ will be taken, for simplicity, of type $~^{\diamond }\mathbf{g}_{\alpha \beta }=\eta _{\alpha \beta}+h_{\alpha \beta }$ for perturbative models with flat background $\eta _{\alpha \beta }.$
Such solutions are with breaking of Lorentz symmetry and, for well stated  conditions,  result into
effective models with covariant renormalization. Using frame transforms $\mathbf{g}%
_{\alpha \beta }=e_{~\alpha }^{\alpha ^{\prime }}e_{~\beta }^{\beta ^{\prime
}}\ ^{\diamond }\mathbf{g}_{\alpha ^{\prime }\beta ^{\prime }},$%
\footnote{%
we label local coordinates and $u^{\alpha }=(x^{i},y^{a})\rightarrow
u^{\alpha ^{\prime }}=(x^{i^{\prime }}(u^{\alpha }),y^{a^{\prime
}}(u^{\alpha })),$ for indices $i,j,k,...=1,2$ and $a,b,c,..=3,4$ (in brief,
we write $u=(x,y),u^{\prime }=(x^{\prime },y^{\prime }));$ the Einstein rule
on summation on repeating ''up-low'' indices will be applied if a contrary
statement will be not emphasized} we shall search for classical and quantum perturbative solutions  for the theories determined by $\ ^{[1]}\mathcal{L}$,   or $\  ^{[2]}\mathcal{L}$.

In this paper, we show how to construct off--diagonal solutions in GR (in general, with nontrivial effective anisotropic polarizations of cosmological
constants), when in a theory for $\ ^{[1]}S$ there are modelled effects with broken Lorenz invariance, non--standard effective anistoropic fluid coupling and behavior of the polarized propagator in the ultraviolet/infrared region. Such effects are  derived to be similar to
those for some (super-) renormalizable theories $\ ^{[3]}S$ and/or $\ ^{[4]}S.$ The solutions for  $\ ^{[2]}S$ will be used as "bridges"  between generating functions determining certain classes of generic off--diagonal Einstein manifolds and effective models with anisotropies and resulting violations for Lorentz symmetry. We state the conditions when the Einstein eqs transform nonlinearly, by imposing corresponding classes of nonholonomic constraints, to necessary types of effective systems of partial differential eqs (PDE) with parametric dependence of solutions which under quantization ''survive'' and stabilize in some rescaled/anisotropic and renormalized forms. Finally, we shall generalize the constructions for exact and approximate anisotropic solutions for generic off--diagonal EYMH systems and their covariant renormalizable models.

\section{Decoupling property and formal integration of Einstein eqs} We briefly review the anholonomic deformation method of constructing
off--diagonal exact solutions in the GR theory and modifications
 \cite{ijgmm2,vexol}. Any spacetime metric can be parametrized in a
form
{\small
\begin{eqnarray}
\mathbf{g} &=&\underline{g}_{\alpha \beta }(u)du^{\alpha }\otimes du^{\beta
}=g_{\alpha }(u)\mathbf{e}^{\alpha }\otimes \mathbf{e}^{\alpha } \notag \\ &=& g_{ij}\
e^{i}\otimes e^{j}+\ g_{ab}\ \mathbf{e}^{a}\otimes \mathbf{e}^{b},
\label{dm} \\
&&\underline{g}_{\alpha \beta }=\left[
\begin{array}{cc}
g_{ij}+N_{i}^{a}N_{j}^{b}g_{ab} & N_{j}^{e}g_{ae} \\
N_{i}^{e}g_{be} & g_{ab}%
\end{array}%
\right] ,  \label{ansatz} \\
\mathbf{e}_{\alpha } &=&[\mathbf{e}_{i}=\partial /\partial x^{i}-
N_{i}^{b}(u)\partial _{b}, e_{a}=\partial _{a}=\partial
/\partial y^{a}],  \label{nder} \\
\mathbf{e}^{\beta } &=&[e^{j}=dx^{j},\ \mathbf{e}^{b}=dy^{b}+\
N_{i}^{b}(u)dx^{i}].  \label{ndif}
\end{eqnarray}%
}
We associate the coefficients $\mathbf{N}=\{N_{i}^{a}(u)\}$ to a 2+2
splitting of $V$ stated explicitly for the tangent space $\mathbf{N:~}%
TV=hV\oplus vV$ for a non--integrable (nonholonomic) distribution with
conventional horizontal, h, and vertical, v, subspaces.\footnote{%
Such a conventional decomposition can be performed
additionally/alternatively / in parallel to the well known $3+1$ splitting
and Arnowit--Deser--Misner, ADM, formalism. Parametrizations of type \ (\ref%
{ansatz}) are used, for instance, in Kaluza--Kelin gravity (see review \cite%
{overduin}). In this paper, we consider 4--d models with nonlinear
dependencies of $N_{i}^{a}(u^{\beta })$ on all coordinates and do not set
compactification on some ''extra" dimensions.}
The decoupling property for the Einstein eqs and their solutions can
proved for any metric $\mathbf{g}$ (\ref{dm}) and parametrization with
''non--underlined'' (in general, depending on variables $(x^{i},y^{3})$) and
''underlined'' multiples (in general, depending on variables $(x^{i},y^{4})$),
{\small
\begin{eqnarray}
g_{i}&=&g_{i}(x^{k}), g_{a}=\omega ^{2}(x^{i},y^{c})h_{a}(x^{k},y^{3})%
\underline{h}_{a}(x^{k},y^{4}),  \label{paramdcoef}  \\
N_{i}^{3}&=&w_{i}(x^{k},y^{3})+\underline{w}%
_{i}(x^{k},y^{4}),
N_{i}^{4}=n_{i}(x^{k},y^{3})+\underline{n}%
_{i}(x^{k},y^{4}). \notag
\end{eqnarray}%
}
 Such functions of necessary smooth class  have to be defined in a form
generating solutions of Einstein eqs. A conformal $v$--factor $\omega
(x^{i},y^{c})$  may depend on all coordinates. We may
simplify substantially the constructions if we take $\omega =\underline{h}%
_{a}=1$ and $\underline{w}_{i}=\underline{n}_{i}=0$ resulting in generic  off--diagonal metrics
with Killing symmetry on $\partial /\partial y^{4}.$ There will be used
brief denotations for partial derivatives: $a^{\bullet }=\partial a/\partial
x^{1},a^{\prime }=\partial a/\partial x^{2},a^{\ast }=\partial a/\partial
y^{3},a^{\circ }=\partial a/\partial y^{4}.$\footnote{%
In this work, we  restrict our constructions to some
classes of spacetime metrics, which can be generated from a set of
nontrivial data $\left[ g_{i},h_{a},w_{i},n_{i}\right] $ with Killing
symmetry by nonholonomic deformations depending on certain small parameters
resulting into general ''non--Killing'' data $\left[ g_{i},\omega ^{2}h_{a}%
\underline{h}_{a},w_{i}+\underline{w}_{i},n_{i}+\underline{n}_{i}\right] $. In a perturbative approach to QG, we can consider that the contributions  from quadratic terms for products of the coefficients of metrics and/or
connections, of type $\Gamma \cdot \Gamma ,$ are small for a fixed open region $U\subset \mathbf{V}$ endowed with normal coordinates.}

The system of Einstein eqs derived for the first action in (\ref{act1}),
where the Ricci tensor $\widehat{R}_{\alpha \beta }$ is computed for (\ref%
{paramdcoef}), can be transformed equivalently into a system of PDE with h--v--decoupling, see details in  \cite{ijgmmp2,vexol}, 
{\small
\begin{eqnarray}
&&g_{2}^{\bullet \bullet }-\frac{g_{1}^{\bullet }g_{2}^{\bullet }}{2g_{1}}-%
\frac{\left( g_{2}^{\bullet }\right) ^{2}}{2g_{2}} g_{1}^{\prime \prime }-%
\frac{g_{1}^{\prime }g_{2}^{\prime }}{2g_{2}}-\frac{(g_{1}^{\prime })^{2}}{%
2g_{1}}=2g_{1}g_{2} \widehat{\Lambda },   \notag 
 \\
&&\frac{-1}{2h_{3}h_{4}}[h_{4}^{\ast
\ast }-\frac{\left( h_{4}^{\ast }\right) ^{2}}{2h_{4}}-\frac{h_{3}^{\ast
}h_{4}^{\ast }}{2h_{3}}]+  \notag \\
&& \frac{1}{2\underline{h}_{3}\underline{h}_{4}}[%
\underline{h}_{3}^{\circ \circ }-\frac{\left( \underline{h}_{3}^{\circ
}\right) ^{2}}{2\underline{h}_{3}}-\frac{\underline{h}_{3}^{\circ }%
\underline{h}_{4}^{\circ }}{2\underline{h}_{4}}]=-\widehat{\Lambda },
\label{eq2} \\
&&\frac{w_{k}}{2h_{4}}[h_{4}^{\ast \ast }-\frac{\left(
h_{4}^{\ast }\right) ^{2}}{2h_{4}}-\frac{h_{3}^{\ast }h_{4}^{\ast }}{2h_{3}}%
]+\frac{h_{4}^{\ast }}{4h_{4}}\left( \frac{\partial _{k}h_{3}}{h_{3}}+\frac{%
\partial _{k}h_{4}}{h_{4}}\right)  \notag  \\
&&  -\frac{\partial _{k}h_{4}^{\ast }}{2h_{4}}
 +\frac{\underline{h}_{3}}{2\underline{h}_{4}}\underline{n}_{k}^{\circ \circ
}+\left( \frac{\underline{h}_{3}}{\underline{h}_{4}}\underline{h}_{4}^{\circ
}-\frac{3}{2}\underline{h}_{3}^{\circ }\right) \frac{\underline{n}%
_{k}^{\circ }}{2\underline{h}_{4}}=0,    \label{eq3} \\
&&\frac{\underline{w}_{k}}{2\underline{h}_{3}}[\underline{h}%
_{3}^{\circ \circ }-\frac{\left( \underline{h}_{3}^{\circ }\right) ^{2}}{2%
\underline{h}_{3}}-\frac{\underline{h}_{3}^{\circ }\underline{h}_{4}^{\circ }%
}{2\underline{h}_{4}}]+\frac{\underline{h}_{3}^{\circ }}{4\underline{h}_{3}}%
\left( \frac{\partial _{k}\underline{h}_{3}}{\underline{h}_{3}}+\frac{%
\partial _{k}\underline{h}_{4}}{\underline{h}_{4}}\right)  \notag \\
&& -\frac{\partial _{k}\underline{h}_{3}^{\circ }}{2\underline{h}_{3}}
+\frac{h_{4}}{2h_{3}}n_{k}^{\ast \ast }+\left( \frac{h_{4}}{h_{3}}%
h_{3}^{\ast }-\frac{3}{2}h_{4}^{\ast }\right) \frac{n_{k}^{\ast }}{2h_{3}}%
=0,   \label{eq4} \\
&&w_{i}^{\ast }=(\partial _{i}-w_{i})\ln |h_{4}|,(\partial
_{k}-w_{k})w_{i}=(\partial _{i}-w_{i})w_{k}, \label{lcconstr1} \\
&&n_{i}^{\ast }=0,\partial _{i}n_{k}=\partial _{k}n_{i}, \notag   \\
&&\underline{w}_{i}^{\circ }=0,\ \partial _{i}\underline{w}_{k}=\partial _{k}%
\underline{w}_{i},\underline{n}_{i}^{\circ }=(\partial _{i}-\underline{n}%
_{i})\ln |\underline{h}_{3}|, \notag \\
&&(\partial _{k}-\underline{n}_{k})\underline{n}%
_{i}=(\partial _{i}-\underline{n}_{i})\underline{n}_{k},   \notag \\
&&\mathbf{e}_{k}\omega = \partial _{k}\omega -\left( w_{i}+\underline{w}%
_{i}\right) \omega ^{\ast }-\left( n_{i}+\underline{n}_{i}\right) \omega
^{\circ }=0.   \label{conf2}
\end{eqnarray}%
} The values $\widehat{R}_{\alpha }^{\alpha }$ and
$\widehat{R}_{ak}$ are equal to the corresponding ones
computed for the Levi--Civita connection if the conditions
(\ref{lcconstr1}) are satisfied. In order to exclude certain
''degenerate'' classes of solutions of above nonlinear
systems of PDEs, we can impose the conditions
$h_{4}^{\ast }\neq 0$ and $\underline{h}%
_{3}^{\circ }\neq 0$. Such conditions can be satisfied always if
corresponding frame/coordinate transforms to necessary ansatz are considered.

At the next step, we show that we can construct in general form a class of
exact solutions with generic off--diagonal metrics  determined by
coefficients with one Killing symmetry, $\left[ g_{i},h_{a},w_{i},n_{i}%
\right] $ with $h_{4}^{\ast }\neq 0.$ For
\begin{equation}
\phi (x^{k},y^{3})=\ln \left| h_{4}^{\ast }/\sqrt{|h_{3}h_{4}|}\right| ,\ \
\alpha _{i}=h_{4}^{\ast }\partial _{i}\phi ,\ \beta =h_{4}^{\ast }\phi
^{\ast }, \label{auxfunct}
\end{equation}%
 we can write respectively the eqs (\ref{eq2}), (\ref{eq3}) in the forms,
\begin{eqnarray}
\phi ^{\ast }h_{4}^{\ast } &=&2h_{3}h_{4}\widehat{\Lambda },\
 \phi ^{\ast }\neq 0,     \label{eq2f}
\\
\beta w_{i}+\alpha _{i} &=&0.  \label{eq3f}
\end{eqnarray}%
For (\ref{eq4}), we must take any trivial solution given by a function $%
n_{i}=$ $n_{i}(x^{k})$ satisfying the conditions $\partial
_{i}n_{j}=\partial _{j}n_{i}$ in order to solve the constraints (\ref%
{lcconstr1}). Using coefficients (\ref{auxfunct}) with $\alpha _{i}\neq 0$
and $\beta \neq 0,$ the solution of the above system of Einstein eqs
with arbitrary off--diagonal coefficients for one--Killing symmetry can
be expressed in the form determined by generating functions $\psi (x^{k}),\phi
(x^{k},y^{3}),$ $\phi ^{\ast }\neq 0,n_{i}(x^{k})$ and $\underline{h}%
_{4}(x^{k},y^{4}),$ and integration function $~^{0}\phi (x^{k})$ following recurrent formulas and conditions,
\begin{equation}
h_{3}=\pm \frac{(\phi ^{\ast })^{2}}{4\widehat{\Lambda }},~h_{4}=\mp \frac{1%
}{4\widehat{\Lambda }}e^{2[\phi -~^{0}\phi ]},w_{i}=-\frac{\partial _{i}\phi
}{\phi ^{\ast }}.  \label{solut1}
\end{equation}%
In the above formulas, we should take respective values
$\epsilon _{i}=\pm 1$ and $\pm $ in (\ref{solut1}) in order to fix a
necessary spacetime signature. The generating/ integration functions
may depend also on arbitrary finite sets of
parameters $\theta =(\ ^{1}\theta ,\ ^{2}\theta ,...)$ at we
found in Ref. \cite{ijgmmp2,vexsol}.
In general, such parametric functions are of type
$\psi (x^{k},\theta )$, $\phi (x^{k},y^{3},\theta )$
etc and their explicit form have to be defined from
certain boundary/asymptotic conditions and/or
experimental data. In some limits, the resulting
solutions may describe configurations with
effective broken Lorentz symmetry, anisotropies,
deformed symmetries etc. For simplicity, we shall
not write in explicit form the parametric dependence
of values if that will not result in ambiguities.
Re--scaling the generating function $\phi $ in
such a form that
$\phi ^{\ast }\rightarrow \phi ^{\ast }\widehat{\Lambda }$
for $\widehat{\Lambda }\neq 0,$ we can generate in a
similar form
solutions with Killing symmetry on $\partial /\partial y^{3}$
with data
$\left[ g_{i},\underline{h}_{a},\underline{w}_{i},\underline{n}_{i}\right] $
for ''rescaled'' generating/integration functions $\underline{\phi }%
(x^{k},y^{4}),\underline{\phi }^{\circ }\neq 0,\underline{w}_{i}(x^{k})$, $%
~^{0}\underline{\phi }(x^{k}),$ when $\underline{h}_{3}=\mp e^{2[\underline{%
\phi }-~^{0}\underline{\phi }]},\underline{h}_{4}=\pm (\underline{\phi }%
^{\circ })^{2},\underline{n}_{i}=-\partial _{i}\underline{\phi }/\underline{%
\phi }^{\circ }$.

We conclude that the Einstein eqs $\widehat{R}_{~\beta }^{\alpha
}=\delta _{~\beta }^{\alpha }\widehat{\Lambda },$ where $\widehat{R}_{~\beta
}^{\alpha }$ is the Ricci tensor and $\widehat{\Lambda }\neq 0$, decouple in
very general forms with respect to N--adapted frames (\ref{nder}) and (\ref%
{ndif}), which allows us to express their solutions (up to corresponding
frame transforms) in a form (\ref{paramdcoef}),
{\small
\begin{eqnarray}
\mathbf{g} &=&\epsilon _{i}e^{\psi (x^{k})}dx^{i}\otimes dx^{i}+
\omega ^{2}
[\epsilon _{3}(\phi ^{\ast })^{2}e^{2(\underline{\phi }-~^{0}\underline{%
\phi })}\mathbf{e}^{3}\otimes \mathbf{e}^{3}+  \notag \\
&& \epsilon _{4}(\underline{\phi }%
^{\circ })^{2}e^{2(\phi -~^{0}\phi )}\mathbf{e}^{4}\otimes
\mathbf{e}^{4}], \label{gensol} \\
\mathbf{e}^{3} &=&dy^{3}+(\underline{w}%
_{i} -\partial _{i}\phi /\phi ^{\ast }) dx^{i},
 \mathbf{e}^{4}=dy^{4}+(n_{i}-\partial _{i}\underline{\phi }/%
\underline{\phi }^{\circ}) dx^{i}, \notag
\end{eqnarray}%
}
 where $\epsilon _{\beta }=\pm 1$ fix a corresponding signature and
 $\omega $ is subjected to constraints of type (\ref{conf2}), $$\mathbf{e}_{k}\omega
=\partial _{k}\omega +\left( \partial _{i}\phi /\phi ^{\ast }-\underline{w}%
_{i}\right) \omega ^{\ast }+\left( -n_{i}+\partial _{i}\underline{\phi }/%
\underline{\phi }^{\circ }\right) \omega ^{\circ }=0.$$ Such classes of
metrics are generic off--diagonal and defined by corresponding sets of
generating/ integration functions and parameters. Other classes of solutions
with $h_{4}^{\ast }=0,$ or $\underline{h}_{3}^{\circ }=0,$ and/or $\widehat{%
\Lambda }=0$ consist some special cases analyzed in \cite{ijgmmp2,vexol}.
Physically important solutions for
black holes/ellipsoids, with singularities and horizons, can be constructed  for
corresponding classes of coefficients of metrics, generating and integration functions etc.

\section{Solutions mimicking non--standard perfect fluid coupling in flat
backgrounds}
 The scalar curvature $\ ^{s}\widehat{R}:=\mathbf{g}^{\alpha
\beta }\widehat{\mathbf{R}}_{4k}$ derived from the Ricci tensors computed for ansatz $\mathbf{g}$  (\ref{paramdcoef}) with
Killing symmetry on $\partial /\partial y^{4}$ and nontrivial  $\left[
g_{i},h_{a},w_{i},n_{i}\right] $ is computed\footnote{%
for simplicity, hereafter, we shall use such classes of off--diagonal
solutions even technically it will be always possible to extend the
constructions to general ones (\ref{gensol})} $\ ^{s}\widehat{R}=-2\widehat{%
\Lambda }-\frac{\phi ^{\ast }h_{4}^{\ast }}{h_{3}h_{4}}=-4\widehat{\Lambda }$%
. Let us state the conditions for generating functions when a metric $%
\mathbf{g}_{\alpha \beta }$ is defined as a solution of field eqs in
theories for $\ ^{[1]}\mathcal{L}$ and/or $~^{[2]}\mathcal{L}$ is also
equivalent to a solution derived for $\ ^{[3]}\mathcal{L}.$ Considering that
$\widetilde{L}$ is a matter Lagrangian density depending only on the metric
tensor components $\mathbf{g}_{\alpha \beta }$ but not on its derivatives,
for the energy--momentum tensor $\widetilde{T}_{\mu \nu }=-\frac{2}{\sqrt{|%
\mathbf{g}|}}\frac{\delta (\sqrt{|\mathbf{g}|}\widetilde{L})}{\delta \mathbf{%
g}^{\mu \nu }}=\mathbf{g}_{\mu \nu }\widetilde{L}-2\frac{\partial \widetilde{%
L}}{\partial \mathbf{g}^{\mu \nu }},$ we can construct sources with
nontrivial\footnote{%
we may consider $\widetilde{L}=\left(
g_{1}^{-1}+g_{2}^{-1}\right) P_{1}+\left( g_{3}^{-1}+g_{4}^{-1}\right) P_{2}$ and
$\left( g_{1}^{-1}+g_{2}^{-1}-2\right) P_{1}+\left(
g_{3}^{-1}+g_{4}^{-1}\right) P_{2} = \widehat{\Lambda }$, \\
$\left( g_{1}^{-1}+g_{2}^{-1}\right) P_{1}+\left(
g_{3}^{-1}+g_{4}^{-1}-2\right) P_{2} =\Upsilon$.
}
$T_{1}^{1}=T_{2}^{2}=\Upsilon (x^{i},y^{3})$ and $T_{3}^{3}=T_{4}^{4}=%
\widehat{\Lambda }$. For the field eqs derived from $~^{[2]}\mathcal{L}%
=~^{s}\widetilde{R}+\widetilde{L},$ for the same $\mathbf{g}$ and $~^{s}%
\widetilde{R}=~^{s}\widehat{R},$ with respect to N--adapted frames, the first eq from (\ref{auxfunct}) and (\ref{eq2f}) modify respectively as
\begin{equation}
\sqrt{|h_{3}h_{4}|}=h_{4}^{\ast }e^{-\widetilde{\phi }}\mbox{ and }%
\widetilde{\phi }^{\ast }h_{4}^{\ast }=2h_{3}h_{4}\Upsilon  \label{aux6}
\end{equation}%
for a new generating function $\widetilde{\phi }(x^{i},y^{3})$ and  $%
\Upsilon (x^{i},y^{3}).$

The theories for $~^{[1]}\mathcal{L}$ $\ $and $~^{[2]}\mathcal{L}$ are
equivalent if their generating functions and sources are related as $%
|\Upsilon |^{-1}\left( e^{2\widetilde{\phi }}\right) ^{\ast }=\widehat{%
\Lambda }^{-1}\left( e^{2\phi }\right) ^{\ast }$. Using this formula and
solutions (\ref{solut1}), we can compute the functional
dependencies
 $h_{4}[\phi ,\widehat{\Lambda }] =\pm (4\widehat{\Lambda })^{-1} (e^{2\phi
}-e^{2~^{0}\phi }) =h_{4}[\widetilde{\phi },\Upsilon ],$
$ h_{3}[\widetilde{\phi },\Upsilon ] = \pm 4 [(\sqrt{|h_{4}|}) ^{\ast }] ^{2}e^{-2\widetilde{%
\phi }},$
when $\widehat{\Lambda }e^{2\widetilde{\phi }}=e^{2~^{0}\widetilde{\phi }%
}\int dy^{3}|\Upsilon |\left( e^{2\phi }\right) ^{\ast }$ (we can use the
inverse formula, $e^{2\phi }-e^{2~^{0}\phi }=\widehat{\Lambda }\int
dy^{3}|\Upsilon |^{-1} (e^{2\widetilde{\phi }}) ^{\ast }$) with $~^{0}%
\widetilde{\phi }(x^{i})$ and integration functions $~^{0}\phi (x^{i})$.

At the next step, we state the condition when a source $\Upsilon $
transforms a theory for $~^{[2]}\mathcal{L}$ into a model for $~^{[3]}%
\mathcal{L}.$ With respect to coordinate frames and for a flat background
metric $\eta _{\alpha \beta },$ we consider a generic off--diagonal metric $%
~^{\diamond }\mathbf{g}_{\alpha \beta }=\eta _{\alpha \beta }+h_{\alpha
\beta }(x^{i},t),$ where $y^{3}=t$ is the timelike coordinate, with chosen
''gauge'' conditions $h_{tt}=h_{t\widehat{i}}=h_{\widehat{i}t}=0,$ for $%
\widehat{i},\widehat{j},=1,2,4$ (on a manifold $\mathbf{V},$ we can consider a ''double'' splitting $(3+1)$ and $(2+2)$). The corresponding Ricci tensor and scalar curvature  are
{\small
\begin{eqnarray*}
~^{\diamond }R_{\widehat{i}\widehat{j}} &=&\frac{1}{2}(h_{\widehat{i}%
\widehat{j}}^{\ast \ast }+\partial _{\widehat{i}}\partial ^{\widehat{k}}h_{%
\widehat{j}\widehat{k}}+\partial _{\widehat{j}}\partial ^{\widehat{k}}h_{%
\widehat{i}\widehat{k}}-\partial _{\widehat{k}}\partial ^{\widehat{k}}h_{%
\widehat{i}\widehat{j}}),\\  \ ^{\diamond }R_{33}&=&-\frac{1}{2}\delta ^{\widehat{i}%
\widehat{j}}h_{\widehat{i}\widehat{j}}^{\ast \ast };
\ ^{\diamond }R = \delta ^{\widehat{i}\widehat{j}}(h_{\widehat{i}\widehat{j}%
}^{\ast \ast }-\partial _{\widehat{k}}\partial ^{\widehat{k}}h_{\widehat{i}%
\widehat{j}})+\partial ^{\widehat{i}}\partial ^{\widehat{j}}h_{\widehat{i}%
\widehat{j}}.
\end{eqnarray*}%
}
We chose a generating function $\widetilde{\phi }(x^{i},t),\widetilde{\phi }%
^{\ast }\neq 0,$ when \newline
 $\ ^{[1]}\mathcal{L}= \ ^{s}R-\widehat{\Lambda }=-3\widehat{\Lambda }-\frac{%
\phi ^{\ast }h_{4}^{\ast }}{h_{3}h_{4}}=~^{[2]}\mathcal{L}=~^{s}\widetilde{R}%
+\widetilde{L} = \ ^{[3]}\mathcal{L}=~^{\diamond }R+~^{\diamond }L,$  with $~^{\diamond }L$ taken for
$\ ^{\diamond }\mathbf{g}_{\alpha \beta }$ and an effective non--standard coupling with a fluid configuration, is
induced by $\phi $ for a generic off--diagonal solution of the system (\ref%
{eq2f}) and (\ref{eq3f}). Einstein manifolds encoding fluid like
configurations are generated if $\widetilde{\phi }$ contains parameters
$\alpha ,\beta ,\rho , \varpi $ introduced into eqs%
\begin{eqnarray}
\frac{\widetilde{\phi }^{\ast }h_{4}^{\ast }}{2h_{3}h_{4}} &=&
-\ ^{\diamond }L=\alpha \rho ^{2}\{[\beta (3\varpi -1)+\frac{%
\varpi -1}{2}]\delta ^{\widehat{i}\widehat{j}}h_{\widehat{i}\widehat{j}%
}^{\ast \ast } +  \notag \\
&& (\varpi +3\varpi \beta -\beta )(\partial ^{\widehat{i}}\partial ^{%
\widehat{j}}h_{\widehat{i}\widehat{j}}-\partial _{\widehat{k}}\partial ^{%
\widehat{k}}h_{\widehat{i}\widehat{j}})\}^{2}.  \label{ngenf}
\end{eqnarray}%
In above formulas, we wrote $\ ^{\diamond }\Upsilon =-~^{\diamond }L$ in order to emphasize that such a source is determined for a metric
$~^{\diamond }\mathbf{g}_{\alpha \beta }$ and Lagrangian $~^{\diamond }L.$
Using this expression and formulas (\ref{solut1}) redefined for (\ref{aux6}),
we find recurrently%
\begin{equation*}
h_{4}=\pm \frac{1}{4}\int dy^{3}|~^{\diamond }\Upsilon |^{-1}(e^{2\widetilde{%
\phi }})^{\ast },\  h_{3}=\pm 4\left[ \left( \sqrt{|h_{4}|}\right)
^{\ast }\right] ^{2}e^{-2\widetilde{\phi }},
\end{equation*}%
for $\widehat{\Lambda }e^{2\widetilde{\phi }}=e^{2~^{0}\widetilde{\phi }%
}\int dy^{3}|~^{\diamond }\Upsilon |\left( e^{2\phi }\right) ^{\ast }$ and $%
w_{i}=-\frac{\partial _{i}\phi }{\phi ^{\ast }}$ used for a  $%
~\ ^{[1]}\mathcal{L}$--theory.

In the limit $\beta \rightarrow (1-\varpi )/2(3\varpi -1)$,
we can express $~^{\diamond }\Upsilon =\alpha \frac{\rho ^{2}}{4}(\varpi
+1)^{2}[(\partial ^{\widehat{i}}\partial ^{\widehat{j}}h_{\widehat{i}%
\widehat{j}}-\partial _{\widehat{k}}\partial ^{\widehat{k}}h_{\widehat{i}%
\widehat{j}})]^{2}$ and generate a class of  Einstein manifolds
{\small
\begin{eqnarray}
\mathbf{g} &=&\epsilon _{i}e^{\psi (x^{k})}dx^{i}\otimes dx^{i}+\epsilon _{3}%
\left[  (\sqrt{|\int dy^{3}|~^{\diamond }\Upsilon |^{-1}(e^{2\widetilde{%
\phi }})^{\ast }|}) ^{\ast }\right] ^{2}  \notag \\ &&
e^{-2\widetilde{\phi }}\mathbf{e}^{3}\otimes \mathbf{e}^{3} +
\epsilon _{4}\int dy^{3}|4\ ^{\diamond }\Upsilon |^{-1}(e^{2\widetilde{%
\phi }})^{\ast }\mathbf{e}^{4}\otimes \mathbf{e}^{4},  \label{sol2} \\
\mathbf{e}^{3} &=&dy^{3}-(\partial _{i}\phi /\phi ^{\ast })dx^{i},\ \mathbf{e%
}^{4}=dy^{4}+n_{i}dx^{i},  \notag
\end{eqnarray}%
}
Such manifolds are with Killing symmetry on $\partial /\partial y^{4}$ and
broken Lorentz invariance because \ the\ source $~^{\diamond }\Upsilon $
does not contain the derivative with respect to $\partial /\partial
y^{3}=\partial _{t}$. Non--Killing configurations of type (\ref{gensol}) can
be constructed for sources $~^{\diamond }\Upsilon (x^{i},y^{3})+~^{\diamond }%
\underline{\Upsilon }~(x^{i},y^{4})$ and nontrivial factors $\omega
(x^{i},y^{a}),$ as more general classes of solutions.

The fact that such locally anisotropic spacetimes are not Lorentz invariant
is not surprising. We have a similar case for the Schwarz\-schild - de
Sitter solutions which are diffeomorphysm invariant and with broken Lorentz
symmetry. For the class of solutions (\ref{sol2}), in the ultraviolet region
where momentum $\mathbf{k}$ is large, the second term for the equivalent
theory $\ ^{[3]}\mathcal{L}$ gives the propagator $|\mathbf{k|}^{-4}.$  The  longitudinal modes do not
propagate being allowed propagation of the transverse one. Such a behavior
is similar to that in a theory with non--standard coupling of gravity with
perfect fluid when the energy--momentum tensor $T_{\widehat{i}\widehat{j}%
}=p\delta _{\widehat{i}\widehat{j}}=\varpi \rho \delta _{\widehat{i}\widehat{%
j}}$ and $T_{33}=\rho $ (if we treat $p,\rho $ and $\varpi $ as standard
fluid parameters and eq of state) in the flat background is computed
for $~^{\diamond }L=-\alpha (T^{\alpha \beta }~^{\diamond }R_{\alpha \beta
}+\beta T_{\alpha }^{\alpha }~^{\diamond }R_{\beta }^{\beta })^{2}.$

Considering off--diagonal solutions for theories with $\ ^{[1]}\mathcal{L}$
and/or $\ ^{[2]}\mathcal{L}$ derived for generating functions and sources of
type (\ref{ngenf}), we obtain a situation when generic off--diagonal
interactions in gravity induce a kind of spontaneous violation of symmetry
which is typical in quantum field theories. Using such nonholonomic
configurations, we model a vacuum gravitational aether via analogous
coupling with non--standard fluid which breaks the Lorenz symmetry for an
effective equivalent theory $\ ^{[3]}\mathcal{L}.$ This allows us to
elaborate a (power--counting) renormalizable model of QG.\footnote{In this letter, we do not prove explicitly the  renormalizability  of off--diagonal solutions but show that they can be associated/ related to certain "renormalizable" quantum models studied by other authors.}

\section{Effective renormalizable EYMH configurations and modified gravity}
The  constructions with non--standard fluid coupling, effective
renromalizability etc performed after (\ref{aux6}) were provided, for simplicity, for
a flat background. Nevertheless, the approach can be extended to curved
backgrounds and generic off--diagonal gravitational--field interactions.
Using the principle of relativity, we can work in a local Lorenz frame when,
for instance, the effective fluid does not flow. This way we preserve
unitarity and axioms of GR working with general class of solutions for theories $\ ^{[1]}\mathcal{L}$ and/or $\ ^{[2]}\mathcal{L}$. Via
respective parametric dependence and nonholonomic transforms we mimic some
models $\ ^{[3]}\mathcal{L}$ with anisotropic coupling. More than that, even the value $~^{\diamond }\Upsilon $ can be fixed to  not contain
time derivatives, it will get such terms in  arbitrary frames of reference.
As we mentioned in  \cite{ijgmmp2,vexol}, the solutions of type (\ref{sol2}) and (%
\ref{gensol}) in the diagonal spherical symmetry limit (here we include the
condition $T^{\alpha \beta }=0) $ contain the
Schwarzschild and Kerr black hole/ellipsoid metrics. The analyzed models
with $\Upsilon =~^{\diamond }\Upsilon =-~^{\diamond }L$ (\ref{ngenf}) result
in $z=2$ Ho\v{r}ava--Lifshitz theories.

Let us show how in the scheme of Lagrangian densities (\ref{act1}) we
can include $z=3$ theories which allows us to generate ultra--violet power
counting renormalizable 3+1 and/or 2+2 quantum models. Instead of  $%
~^{\diamond }L,$ we can take more general sources and generating functions, $-\  _{n}^{\diamond }L=$
{\small
\begin{equation}
\alpha \{(T^{\mu \nu }\ ^{\diamond }\nabla _{\mu}\ ^{\diamond }\nabla _{\nu }+\gamma T_{\alpha }^{\alpha }
 \  ^{\diamond }\nabla ^{\beta }\ ^{\diamond }\nabla _{\beta})^{n}
   (T^{\alpha \beta }\ ^{\diamond }R_{\alpha \beta }+\beta T_{\alpha
}^{\alpha }\ ^{\diamond }R_{\beta }^{\beta })\}^{2},  \label{aux7}
\end{equation}%
}
where $n$ and $\gamma $ are constants. Introducing $\Upsilon
=~-~_{n}^{\diamond }L$ in formulas (\ref{ngenf}) in order to state a
different class of generating functions $~^{n}\widetilde{\phi }$ from $\frac{%
~^{n}\widetilde{\phi }^{\ast }h_{4}^{\ast }}{2h_{3}h_{4}}=-~_{n}^{\diamond }L
$ and construct off--diagonal solutions of type (\ref{sol2}) for data $%
\widetilde{\phi }\rightarrow ~^{n}\widetilde{\phi },\phi \rightarrow
~^{n}\phi $ and $~^{\diamond }\Upsilon \rightarrow -~_{n}^{\diamond }L.$ \
In general, we can consider noninteger values for $n,$ for instance, $%
n=1/2,3/2$ etc. If $n$ is negative, we can generate non--local theories
determined by a respective off--diagonal polarization of vacuum and
cosmological constant in GR (the question of ''renormalizability'' is more
complex for such cases). Following analysis provided in
\cite{odints1}
the analogous $\ ^{[3]}\mathcal{L}$ and $\ ^{[4]}\mathcal{%
L}$ models derived for (\ref{aux7}) are renormalizable if $n=1$ and
super--renormalizable for $n=2.$ The values induced by a nontrivial
source/Lagrangian density $~_{n}^{\diamond }L$ contains higher derivative
terms and effectively break the Lorenz symmetry for high energies, i.e. in
UV region. In the IR, we positively get the usual Einstein gravity.
Nevertheless, in our approach, GR is not only a limit from certain
modifications with non-standard coupling of type (\ref{ngenf}) and/or (\ref%
{aux7}) but a model with possible ''branches" of complexity, anisotropies, inhomogeneities and Lorentz violations depending on
parameters and generating functions. Certain families of
solutions are (super) renormalizable because of   off--diagonal
nonlinear interactions of gravitational and effective matter fields.

It is important to analyze two special cases for sources $~_{n}^{\diamond }L,
$ for instance, in the fluid approximation for matter. We model two theories
$\ ^{[1]}\mathcal{L}$ and $\ ^{[3]}\mathcal{L}$ with cosmological constant
when both $\varpi =-1$ and $\ \widehat{\Lambda }\neq 0.$ This corresponds to
diagonal \ solutions with $T^{\alpha \beta }\ ^{\diamond }R_{\alpha
\beta }+\beta T_{\alpha }^{\alpha }\ ^{\diamond }R_{\beta }^{\beta }=0$
when $1/|k|^{4}$ can not be obtained for the propagator. To correct the
model, we have to introduce additional off--diagonal terms. The case $\varpi
=1/3$ corresponds to the radiation of conformal matter with divergent $\beta
.$ This also does no provide ''good'' solutions if other ''anisotropic''
contributions are not considered.

The approach with generic off--diagonal solutions modelling effective
covariant renormalizable theories can be extended for gravitational
interactions with a non--Abelian SU(2) gauge field $\mathbf{A}=\mathbf{A}%
_{\mu }\mathbf{e}^{\mu }$ coupled to a triplet Higgs field $\Phi ,$ see
details in  \cite{ijgmmp2,vexol}.  In order to prove the decoupling property, we have
to use a covariant operator $\widehat{\mathbf{D}}$ which is adapted to the
N--splitting. It is not possible to ''see'' a general splitting of Einstein
and matter field eqs if we do not consider $2+2$ spacetime
decompositions with such non--integrable distributions. The linear
connection $\widehat{\mathbf{D}}$ is equivalent to the Levi--Civita
connection $\nabla $ if the conditions (\ref{lcconstr1}) are satisfied and
all computations are performed with respect N--adapted bases. In terms of $%
\widehat{\mathbf{D}}$, such nonholonomic interactions are described%
\begin{eqnarray}
\widehat{\mathbf{R}}_{\ \beta \delta }-\frac{1}{2}\mathbf{g}_{\beta \delta
}\ ^{s}\widehat{R} &=&8\pi G\left( \ ^{H}T_{\beta \delta }+\ ^{YM}T_{\beta \delta }\right) ,  \label{ym1} \\
 D_{\mu }(\sqrt{|g|}F^{\mu \nu }) &=&\frac{1}{2}
 i e (\sqrt{|g|})\ [\Phi ,D^{\nu }\Phi ],  \label{heq2} \\
 D_{\mu }(\sqrt{|g|}\Phi ) &=&\lambda
(\sqrt{|g|})\  (\Phi _{\lbrack
0]}^{2}-\Phi ^{2})\Phi ,  \label{heq3}
\end{eqnarray}%
where the source of the Einstein eqs is
{\small
\begin{eqnarray}
\ ^{H}T_{\beta \delta }=Tr[\frac{1}{4} (\nabla _{\delta }\Phi \ \nabla
_{\beta }\Phi +\nabla _{\beta}\Phi \ \nabla _{\delta }\Phi )
 -\frac{1}{4} \mathbf{g}_{\beta \delta}\nabla _{\alpha}
\Phi \ \nabla ^{\alpha}\Phi ]
 \label{source1} \\
-\mathbf{g}_{\beta \delta }\mathcal{V}(\Phi),\
\ ^{YM}T_{\beta \delta }=2Tr  (\mathbf{g}^{\mu \nu }F_{\beta \mu
}F_{\delta \nu }-\frac{1}{4}\mathbf{g}_{\beta \delta }F_{\mu \nu }F^{\mu \nu
}).  \label{source2}
\end{eqnarray}%
}
 The curvature of gauge field $\mathbf{A}_{\delta }$ is $F_{\beta \mu }=%
\mathbf{e}_{\beta }\mathbf{A}_{\mu }-\mathbf{e}_{\mu }\mathbf{A}_{\beta }+ie[%
\mathbf{A}_{\beta },\mathbf{A}_{\mu }],$  where $e$ is the coupling constant,
$i^{2}=-1,$ and $[\cdot ,\cdot ]$ is used for the commutator.  $\Phi
_{\lbrack 0]}$ in (\ref{heq3}) is the vacuum expectation of the Higgs field
which determines the mass $\ ^{H}M=\sqrt{\lambda }\eta ,$ when $\lambda $ is
the \ constant of scalar field self--interaction with potential $\mathcal{V}%
(\Phi )=\frac{1}{4}\lambda Tr(\Phi _{\lbrack 0]}^{2}-\Phi ^{2})^{2}.$ The
gravitational constant $G$ defines the Plank mass $M_{Pl}=1/\sqrt{G};$ the
is a nontrivial mass of gauge boson, $\ ^{W}M=ev.$

A class of diagonal solutions of the system (\ref{ym1})--(\ref{heq3}), see Ref. \cite{br3}, is given by metric ansatz $\ ^{\circ }\mathbf{g} =$
{\small
\begin{eqnarray}
\ ^{\circ }g_{i}(x^{1})dx^{i}\otimes dx^{i}+\
^{\circ }h_{a}(x^{1},x^{2})dy^{a}\otimes dy^{a} =
  q^{-1}(r)dr\otimes dr \notag \\  + r^{2}d\theta \otimes d\theta +
r^{2}\sin ^{2}\theta
d\varphi \otimes d\varphi -\sigma ^{2}(r)q(r)dt\otimes dt, \label{ansatz1}
\end{eqnarray}%
}
where the coordinates and metric coefficients are parametrized,
respectively, $u^{\alpha }=(x^{1}=r,x^{2}=\theta ,y^{3}=\varphi ,y^{4}=t)$
and $\ ^{\circ }g_{1}=q^{-1}(r),\ ^{\circ }g_{2}=r^{2},\ ^{\circ
}h_{3}=r^{2}\sin ^{2}\theta ,\ ^{\circ }h_{4}=-\sigma ^{2}(r)q(r),$ for $%
q(r)=1-$ $2m(r)/r-\Lambda r^{2}/3,$ where $\Lambda $ is a cosmological constant. The function $m(r)$ is usually interpreted as the total mass--energy within the radius $r$ which for $m(r)=0$ defines an empty de Sitter, $dS,$ space written in a static coordinate system with a
cosmological horizon at $r=r_{c}=\sqrt{3/\Lambda }.$ An ansatz for solution of Yang--Mills (YM) eqs on (\ref{ym1}) on a spherically symmetric background in GR is defined by a single magnetic potential $v(r),$
 $\ ^{\circ }A= \ ^{\circ }A_{2}dx^{2}+\ ^{\circ }A_{3}dy^{3}  = \frac{1}{2e}\left[
v(r)\tau _{1}d\theta +(\cos \theta \ \tau _{3}+v(r)\tau _{2}\sin \theta )\
d\varphi \right]$,
where $\tau _{1},\tau _{2},\tau _{3}$ are Pauli matrices, and for eqs (\ref{heq3}) of the Higgs (H) field is given by $\Phi =\ ^{\circ }\Phi =\chi (r)\tau _{3}.$ The functions $\sigma (r),q(r),v(r),\chi (r)$ can be computed
for data $\left[ \ ^{\circ }\mathbf{g}(r)\mathbf{,}\ ^{\circ }A(r),\ \
^{\circ }\Phi (r)\right] .$ For instance, the diagonal Schwarz\-schild--de
Sitter solution of (\ref{ym1})--(\ref{heq3}) is that determined by data$%
v(r)=\pm 1,\sigma (r)=1,\phi (r)=0,q(r)=1-2M/r-\Lambda r^{2}/3$ defining a
black hole configuration inside a cosmological horizon because $q(r)=0$ has two positive solutions and $M<1/3\sqrt{\Lambda }.$

A ''prime'' diagonal solution $\ ^{\circ }\mathbf{g}$ (\ref{ansatz1}) can be
transformed into ''target''  off--diagonal metrics $\mathbf{g}=
\ ^{\eta }\mathbf{g}, \ ^{\circ }\mathbf{g\rightarrow }\ ^{\eta }\mathbf{g,}$ {\small
\begin{eqnarray}
\ ~^{\eta }\mathbf{g} &=&\eta _{i}\ e^{\psi }dx^{i}\otimes
dx^{i}+\omega ^{2}[\eta _{3} (\phi ^{\ast })^{2}e^{2(\underline{%
\phi }-\ ^{0}\underline{\phi })}\mathbf{e}^{3}\otimes \mathbf{e}^{3}
 + \notag \\
 && \eta _{4}(\underline{\phi }^{\circ })^{2}e^{2(\phi -~^{0}\phi )}\mathbf{e%
}^{4}\otimes \mathbf{e}^{4}], \label{dm1}  \\
\mathbf{e}^{3} &=&dy^{3}+\eta _{i}^{3}(w_{i}+\underline{w}_{i})dx^{i},\
\mathbf{e}^{4}=dy^{3}+\eta _{j}^{4}(n_{j}+\underline{n}_{j})dx^{j},
 \notag
\end{eqnarray}%
}
where the gravitational $\eta $--polarizations $\eta _{i}(x^{k}),\eta
_{b}(x^{k},y^{a})$ and $\eta _{j}^{c}(x^{k},y^{a})$ have to be found from
the condition that such metrics generate solutions of (\ref{ym1})--(\ref%
{heq3}). With respect to N--adapted frames, the gauge fields are deformed
\begin{equation}
A_{\mu }(x^{i},y^{3})=\ ^{\circ }A_{\mu }(x^{1})+\ ^{\eta }A_{\mu
}(x^{i},y^{a}),  \label{ans2a}
\end{equation}%
where $\ ^{\eta }A_{\mu }$ is a function for which
 $F_{\beta \mu }=\ ^{\circ }F_{\beta \mu }(x^{1})+ \ ^{\eta }F_{\beta \mu}(x^{i},y^{a})=s\sqrt{|g|}\varepsilon _{\beta \mu },$
for $s=const$ and $\varepsilon _{\beta \mu }$ being the absolute
antisymmetric tensor. Such a tensor  always solve the eqs
$D_{\mu }(\sqrt{|g|}F^{\mu \nu })=0$,
which always give us the possibility to determine the distortions $\ ^{\eta
}F_{\beta \mu }$, and $\ ^{\eta }A_{\mu }$, for any given
$\ ^{\circ }A_{\mu}$ and/or $\ ^{\circ }F_{\beta \mu }.$ The scalar field is nonholonomically
modified $\ ^{\circ }\Phi (x^{1})\rightarrow \Phi (x^{i},y^{a})=\ ^{\Phi
}\eta (x^{i},y^{a})\ ^{\circ }\Phi (x^{1})$ by  $\ ^{\Phi
}\eta $ is such a way that
 $D_{\mu }\Phi =0$ and $\Phi (x^{i},y^{a})=\pm \Phi _{\lbrack 0]}$.
This nonholonomic configuration of the nonlinear scalar field is not trivial
even with respect to N--adapted frames $\mathcal{V}(\Phi )=0$ and $\
^{H}T_{\beta \delta }=0,$ see (\ref{source1}).\footnote{%
For the ansatz (\ref{dm1}), the Higgs eqs are
 $(\partial /\partial x^{i}-A_{i})\Phi = (w_{i}+\underline{w}_{i})\Phi ^{\ast
}+(n_{i}+\underline{n}_{i})\Phi ^{\circ },$
 $\left( \partial /\partial
y^{3}-A_{3}\right) \Phi = 0,  \left( \partial /\partial y^{4}-A_{4}\right) \Phi =0,$
i. e. a scalar field $\Phi $ modifies the off--diagonal components of the
metric via $w_{i}+\underline{w}_{i}$ and $n_{i}+\underline{n}_{i}$ and
nonholonomic conditions for $A_{\mu }=\ ^{\eta }A_{\mu }.$}

The gauge fields  with  potential $A_{\mu }$ (\ref{ans2a}) modified nonholonomically by $\Phi $  determine exact
solutions for the YMH system (\ref{heq2}) and (\ref{heq3}) for spacetime metrics type (\ref{dm1}). The corresponding stress energy--momentum tensor is computed (see details in sections 3.2 and 6.51 from \cite{lidsey}) $\ ^{YM}T_{\beta }^{\alpha }=-4s^{2}\delta _{\beta }^{\alpha },$ which means
that nonholonomically interacting gauge and Higgs fields, with respect to
N--adapted frames, result in an effective cosmological constant $\
^{s}\lambda =8\pi s^{2}$ which should be added to a respective sources of Einstein eqs. Using above presented constructions, we conclude that
various classes of nonholonomic EYMH systems can be modeled, respectively,
as theories of type$\ ^{[1]}\mathcal{L},~^{[2]}\mathcal{L}$ and $\ ^{[3]}%
\mathcal{L}$ using an effective cosmological constant $\widehat{\Lambda }%
=\ \Lambda +\ ^{s}\lambda $ and off--diagonal ansatz for metrics
(\ref{dm1}) when $\eta $--polarizations are treated as additional
generating functions. Instead of such classes of equivalence of conjectured covariantly renormalizable theories with locally anisotropic interactions,
we can reformulate the results in the language of $F(R)$--gravity of Ho\v{r}ava--Lifshitz type
  \cite{kluson}.
The effective Lagrangian density $\ ^{[4]}\mathcal{L}=F(%
\breve{R})$ can be chosen for $~\breve{R}=R+~^{\diamond }L,$ see (\ref{ngenf}%
), when $z=2,$ or $\breve{R}=R+~_{n}^{\diamond }L,$ see (\ref{aux7}), when $%
z=2n+2.$ EYMH configurations are expected to be covariantly renormalizable when $z\geq 3$ which may be demonstrated   using arguments presented in \cite{vtwo,odints1}
and in this paper and/or other variants of anisotropic   renormalization. A rigorous proof  is possible if we chose generating functions when our off--diagonal metrics induce models with covariant power--counting  renormalization \cite{odints3a,odints3b,odints3c}; such constructions will be provided in our future works.

\section{Concluding Remarks}
 In summary, we have used the possibility to
decouple in very general forms the gravitational field eqs in GR
and construct various classes of exact generic off--diagonal solutions
in order to elaborate covariant (super) renormalizable theories of gravity.
We also briefly considered how EYMH interactions can be encoded into
nonholonomic Einstein manifolds and effective models of interactions with broken Lorenz symmetry and covariant renormalization.

We emphasize that in this work we followed an ''orthodox'' approach
to modeling classical and quantum geometries and physical theories of
interactions keeping the constructions and physical paradigm to be maximally
closed to Einstein gravity. Our opinion  is that
the bulk of experimental data for modern gravity and cosmology can be
explained/predicted using generic off--diagonal solutions in GR and their quantized versions.

  \acknowledgments
  I thank E. Elizalde and  S. Odintsov for hospitality and important critical remarks and references. The research in this paper is partially supported by the Program IDEI, PN-II-ID-PCE-2011-3-0256.


\begin{thebibliography}{99}
  \bibitem{alvarez}
\Name{Alvarez E.}
\REVIEW{Rev. Mod. Phys.}{61}{1989}{561}

\bibitem{buch}
  \Name{Buchbinder I. L., Odintsov S.  \and Shapiro I.}
  \Book{Effective Action in Quantum Gravity}
  \Publ{IOP, London}
  \Year{1992}
 \bibitem{oriti}
 \Editor{Oriti D.}
  \Book{Approaches to Quantum Gravity, Toward a
New Understanding of Space, Time and Matter}
   \Publ{Cambridge Univ. Press, Cambridge}
  \Year{2009}

\bibitem{horava}
\Name{Ho\v rava P.}
\REVIEW{Phys. Rev. D}{79}{2009}{084009}

\bibitem{lumei}
\Name{Lu H., Mei  J. \and  Pope C. N.}
\REVIEW{Phys. Rev. Lett.}{103}{2009}{091301}

\bibitem{moffat}
\Name{Moffat J. W.}
\REVIEW{Class. Quant. Grav.}{27}{2010}{135016}

\bibitem{odints1}
\Name{Nojiri S.,Odintsov S.}
\REVIEW{Phys.Rev.D}{81}{2010}{043001}

\bibitem{ijgmmp2}
\Name{Vacaru S.}
\REVIEW{IJGMMP}
{8}{2011}{9}

\bibitem{vexol}
\Name{Vacaru S.}
{ arXiv: 11082022}{}{}{}

\bibitem{kluson}
 \Name{Carloni S., Chaichian M., Nonjiri S., Odintsov S. D.,  Oksanen M.  \and Tureanu A.}{ arXiv: 1003.3925}{}{}{}

\bibitem{overduin}
\Name{ Overduin  J. M. \and Wesson P. S. }
 \REVIEW{Phys. Rept.}
 {283}{1997}{303}

\bibitem{br3}
\Name{Brihaye Y., Hartmann  B., Radu E.  \and Stelea C.}
 \REVIEW{Nucl. Phys. B}
 {763}{2007}{115}

\bibitem{lidsey}
\Name{Lidsey J. E., Wands D.  \and Copeland E. J.}
 \REVIEW{Phys. Rept.}
 {337}{2000}{343}

\bibitem{vtwo}
 \Name{Vacaru S.}
  \REVIEW{IJGMMP}
  {7}{2010}{713}
     \bibitem{odints3a}
  \Name{Nojiri S. \and Odintsov S. D.}
{arXiv: 1004.3613}{}{}{}
  \bibitem{odints3b}
   \Name{Kluson J., Nojiri S.,Odintsov S.}
{arXiv: 1104.4286}{}{}{}
  \bibitem{odints3c}
   \Name{ Nojiri S.,Odintsov S.}
{arXiv: 1007.4856}{}{}{}
   \end{thebibliography}
\end{document}